**Modal decomposition of a propagating matter wave via electron ptychography**

S. Cao[1], P. Kok[2], P. Li[1], A. M. Maiden[1], J. M. Rodenburg[1]

[1]*Department of Electronic and Electrical Engineering, University of Sheffield, S1 3JD, United Kingdom*

[2]*Department of Physics and Astronomy, University of Sheffield, S3 7RH, United Kingdom*

Abstract:

We employ ptychography, a phase-retrieval imaging technique, to show experimentally for the first time that a partially coherent high-energy matter (electron) wave emanating from an extended source can be decomposed into a set of mutually independent modes of minimal rank. Partial coherence significantly determines the optical transfer properties of an electron microscope and so there has been much work on this subject. However, previous studies have employed forms of interferometry to determine spatial coherence between discrete points in the wavefield. Here we use the density matrix to derive a formal quantum mechanical description of electron ptychography and use it to measure a full description of the spatial coherence of a propagating matter wavefield, at least to the within the fundamental uncertainties of the measurements we can obtain.

# 1. INTRODUCTION

We investigate experimentally the decomposition of a freely-propagating partially coherent electron wave into a set of mutually orthogonal wave function components (modes) which are incoherent with each other. By diagonalising the density matrix, we measure the most compact representation of this mixed state. We use the decomposition to determine explicitly, without making any assumptions, the spatial coherence properties of a field emission electron source, at least within the quantum mechanical uncertainties of the wave we physically measure, which propagates through an aperture of finite lateral extent. Although the principle of the modal decomposition is well known in optics (visible light and X-ray) [1,2] including partial coherence [3, 4], and its relation to the density matrix formalism has been mentioned in the context of X-ray ptychography [5], this is the first time it has been demonstrated experimentally for a matter wave.

The partial coherence of a high-energy electron wave has critical importance in the field of electron microscopy. It defines the transfer properties of the imaging lens in a transmission electron microscope (TEM), determining the extent of the 'information limit' – i.e. the highest spatial frequency that can be transferred through the electron optics, and hence the highest resolution that can be obtained in the image [6,7]. This is particularly important in the field of imaging nanoscale biological structures such as large proteins and viruses because contrast in the image of such weakly-scattering structures must be enhanced by large defocus in the object lens [8], which, in the presence of partial coherence, greatly reduces the width of the information limit and hence limits resolution. In the case of scanning transmission electron microscopy (STEM), the intensity of the focused probe is convolved with the demagnified source intensity, which also imposes a limit in resolution. Perhaps most important of all, the entire field of electron holography depends wholly upon an interference effect that is determined by the coherence properties of the electron wave [9].

In the field of light optics, a common approach for quantifying coherence is to consider the temporal correlation and the underlying statistics of fluctuations of the electric field between pairs of points separated in space. If the source is quasi-monochromatic and spatially localised (as in the experiments described here), many approximations can be employed, greatly abridging the formal treatment. In particular, lateral spatial coherence between points lying within a plane perpendicular to the direction of propagation can be characterised by a single two-dimensional, shift invariant function: the mutual coherence function [10]. For electrons one way to achieve a quantitative measurement of this is to place a wire, which acts as a bi-prism, within the plane of the wavefront, as used in electron holography [9]. As a voltage on the wire is increased, parts of the wavefield either side of it are deflected so that increasingly separated parts of the original wave overlap with one another in a measurement plane downstream of the beam splitter. The degree of coherence is expressed in the decreasing depth of the resulting interference fringes as a function of bi-prism voltage [11]. The incoherent source profile can then be obtained via Fourier transformation of this function (i.e. by using the van Cittert-Zernike theorem [10]). Another method of achieving the interference (this time in reciprocal space) is via convergent beam electron diffraction [12], where a crystal is used as an interferometer. Other less direct methods, e.g. by observing the depth and extent of Fresnel fringes in the image plane, can also be used [13].

Partial coherence can alternatively be modelled as a series of independent modes that are entirely mutually incoherent with respect to one another, but which propagate through the optical system independently as pure coherent states [1,2]. This methodology, which we adopt here, is widely used in the optics community, say for describing modes in a laser source [14]. In quantum mechanics, modes in a matter wave are more naturally – though equivalently – handled by the density matrix formalism [15]. In a particular mode decomposition, the density matrix for the electron wave is not generally diagonal, with off diagonal terms arising from coherence between the modes. However, there is a unique mode decomposition in which the density matrix is diagonal and where the associated eigenmodes are perfectly incoherent and orthogonal with respect to one another (assuming non-degenerate eigenvalues). If the state in each mode is normalised, the eigenvalues of the density matrix are proportional to the probability of finding the system in that mode.

In this paper, we measure the diagonalised density matrix of a freely-propagating mixed state electron wave after it has passed through a confining aperture. We calculate the orthogonal pure states relating to the diagonal terms as they express themselves within the plane of the aperture. This represents the most compact and complete description of the spatial partial coherence in the beam. We demonstrate that as the preparation of the incoherent source is changed (by optically altering its shape) the relative probability of these states change as expected.

The experimental technique we employ is a form of electron ptychography [16-18]. We place a two-dimensional object (a transmission electron microscope test specimen) within the electron beam and measure many diffraction patterns scattered from it as it is moved relative to the aperture. (In fact, the aperture lies in an image plane of the specimen, but the measured data are equivalent.) The distance of the specimen movement between exposures is much less than the size of the aperture, so any one point in the object contributes to many diffraction patterns. By applying the constraints that the illuminating wave and the object function remain constant, it is well known that it is easy to solve for both the object and illumination [19,20]. If the illumination is partially coherent, it is impossible to separate the corresponding intensity at any single detector pixel in any single diffraction pattern into separate modes. However, ptychography makes many measurements, sampling diverse scattering conditions as the object is moved laterally, and since the illumination (and the modes within it) remain constant throughout the process, this gives sufficient data to decompose the illumination modes, as previously shown in the case of photons [5,21]. As an imaging technique that offers wavelength-limited resolution and a very sensitive phase image, electron ptychography is hampered by the fact that electron sources are substantially incoherent: although field emission sources are often described in the literature as fully spatially coherent [16,22], in practice they are not. In fact, the sort of modal decomposition we apply here can be used to mitigate the effects of partial coherence, which can be used to improve any result obtained by electron ptychography [22].

## 2.THEORY

Consider an idealised experimental setup illustrated in Figure 1.a. An incoherent source of electrons of finite size creates a freely propagating wave that impinges upon an aperture. We assume that the distance from the source to the aperture is sufficiently large so that the aperture is effectively lying in the Fraunhofer plane relative to the source. A detector lies downstream of the aperture at a large enough distance so that it lies in the Fraunhofer plane of the aperture.

Even though there are many electrons in the beam, they can be considered non-interacting and are therefore well-described by single particle states. First suppose that the electrons are each in a pure state. We may choose to expand this state in any set of basis functions; for the present discussion we use as an example

$$|\psi_S\rangle = \sum_{lm} d_{lm} |l,m\rangle \,, \tag{1}$$

where $\left\{|l,m\rangle\right\}$ is the Laguerre-Gaussian transverse mode expansion such that $\langle x | l,m \rangle \equiv L_{lm}(x)$ are the Laguerre-Gaussian mode functions. These mode functions are convenient in describing approximately cylindrically symmetric sources. When the state propagates to the aperture plane, it becomes

$$|\psi_A\rangle = \Im|\psi_S\rangle = \sum_{lm} \gamma_{lm} |l,m\rangle, \tag{2}$$

where $\Im$ is the Fourier transform, and $\gamma_{lm}$ are the transformed $d_{lm}$. The aperture $A$ will block the electron beam at certain positions according to the function

$$q(x) = \begin{cases} 1 & if \ x \in A, \\ 0 & if \ x \notin A. \end{cases} \tag{3}$$

The aperture can then be described by the high-rank projector

$$P_A = \sum_x q(x) |x\rangle\langle x|. \tag{4}$$

After the aperture, the state must be renormalized, and becomes

$$|\Psi\rangle = \frac{P_A |\psi_A\rangle}{\sqrt{\|P_A\psi_A\|}} \equiv \eta^{-\frac{1}{2}} P_A |\psi_A\rangle, \tag{5}$$

where $\eta$ indicates the transmittance of the aperture given the state $|\psi_A\rangle$, or how many of the electrons in the state pass through the aperture. We can write $|\Psi\rangle$ in terms of the position states $|x\rangle$ at the aperture:

$$|\Psi\rangle = \eta^{-\frac{1}{2}} \sum_x \sum_{lm} q(x) \gamma_{lm} L_{lm}(x) |x\rangle. \tag{6}$$

In the detector plane the state is modified by another Fourier transform, and becomes

$$|\Psi_D\rangle = \Im(\Psi) = \eta^{-\frac{1}{2}} \sum_x \sum_{lm} q(x) \gamma_{lm} L_{lm}(x) \Im |x\rangle. \tag{7}$$

Finally, the intensity at the position $u$ in the detector plane is

$$I(u) = \left| \langle u | \Psi_D \rangle \right|^2 = \eta^{-1} \left| \sum_{lmx} q(x) \gamma_{lm} L_{lm}(x) \langle u | \Im | x \rangle \right|^2. \tag{8}$$

The matrix elements of the Fourier transform are

$$\langle u|\Im|x\rangle = \frac{1}{N}\exp\left[\frac{2\pi i(xu_x + yu_y)}{N}\right], \tag{9}$$

where $N$ is the number of pixels of the detector. This describes the case of a source of completely coherent electron wave functions.

We now extend the above treatment to take account of the extended nature of the source, which renders the electron beam partially coherent, by considering the case of the mixed states. The density matrix for single electron state for an extended source is a probabilistic mixture of pure states, and can be written as

$$\rho_s = \sum_j p_j \left|\psi_{S,j}\right\rangle\left\langle\psi_{S,j}\right| \qquad with \qquad \left|\psi_{S,j}\right\rangle = \sum_{lm} d_{lm}^{(j)}\left|l,m\right\rangle, \tag{10}$$

and in the aperture plane just before the state becomes

$$\rho_A = \Im\rho_S\Im^{-1}. \tag{11}$$

Applying the aperture projector $P_A$ as before, the state must be renormalized to

$$R_A = \frac{P_A\rho_A P_A}{\|P_A\rho_A P_A\|} = \eta^{-1}\sum_{x,x'} q(x)q(x')\rho_A(x,x')\left|x\right\rangle\left\langle x'\right|, \tag{12}$$

where $\rho_A(x,x') = \left\langle x\right|\rho_A\left|x'\right\rangle$. In the detector plane, we again apply a Fourier transform to $R_A$ to obtain

$$R_D = \Im R_A\Im^{-1} = \eta^{-1}\sum_{x,x'} q(x)q(x')\rho_A(x,x')\Im\left|x\right\rangle\left\langle x'\right|\Im^{-1}. \tag{13}$$

The intensity at position $u$ in the detector plane then becomes

$$I(u) = Tr\left[\left|u\right\rangle\left\langle u\right|R_D\right] = \eta^{-1}\sum_{x,x'} q(x)q(x')\rho_A(x,x')\left\langle u|\Im|x\right\rangle\left\langle x'\right|\Im^{-1}\left|u\right\rangle$$
$$= \frac{1}{\eta N^2}\sum_{x,x'} q(x)q(x')\rho_A(x,x')e^{2\pi i[u_x(x-x')+u_y(y-y')]/N} \tag{14}$$

The matrix element $\rho_A(x, x')$ can be expressed as

$$\rho_A(x, x') = \sum_j \sum_{l_j, m_j} \sum_{l_j', m_j'} p_j \gamma_{l_j m_j}^{(j)} \gamma_{l_j' m_j'}^{(j)*} L_{l_j m_j}(x) L_{l_j' m_j'}^*(x) \ . \tag{15}$$

Note that we can in principle use different transverse mode functions $L(x)$, which will have corresponding different amplitudes $\gamma$ .

Finally, we include an object with complex transmission function $O(x)$ immediately following the aperture. The transmission operator acting on the state can be written as

$$\hat{E}_O(v) = \sum_x O(x - v)|x\rangle\langle x| \qquad with \qquad 0 \le |O(x)| \le 1 \ . \tag{16}$$

Moving the object in the aperture plane along a vector $v_j$ modifies the object function according to

$$O(x) \rightarrow O(x - v_j). \tag{17}$$

We assume that the aperture and object are sufficiently close together that there is negligible propagation between them, and $\left[ P_A, \hat{E}_O(v) \right] = 0$ for all $v$ . The intensity $I_j(u)$ at the position $u$ in the detector plane for the $j^{th}$ object shift is then given by

$$I_j(u) = \frac{1}{\eta_j N^2} \sum_{x, x'} q(x) q(x') O(x - v_j) O^*(x' - v_j) \rho_A(x, x')$$
$$\times \exp\left\{ \frac{2\pi i \left[ u_x(x - x') + u_y(y - y') \right]}{N} \right\} , \tag{18}$$

where

$$\eta_j = Tr\left[ \left| \hat{E}_O(v_j) \right|^2 P_A \rho_A(x, x') \right] \tag{19}$$

is the modified normalisation function due to the shifted object function. In the case of electron scattering through a thin object, the modulus of the transmission function $O(x)$ is often assumed to be unity. The phase of the transmission is also often assumed to be weak (much less than $\pi$), although in practice it is often larger than this. To a good approximation we can therefore write $\hat{E}_O(v_j) = \exp\left[i\varphi(x)\right]$, and as a consequence $\eta_j = \eta$ in this case.

The diffraction patterns in equation (18) are fully specified by the aperture function $q(x)$, the object function $O(x)$ and the state $\rho_A$ whose matrix elements are given by equation (15).

Rather than expanding the transverse mode shape of the electron beam in Laguerre-Gaussian modes, we can expand it into eigen-modes of the density operator. Formally, this can be written as

$$\rho_S = \sum_k s_k \left| m_k \right\rangle \left\langle m_k \right|. \qquad (20)$$

Ptychography can determine the probabilities $s_k$ and the mode functions $m_k(x) = \left\langle x | m_k \right\rangle$, as described below. We assume for simplicity that the $s_k$ are non-degenerate, so that the $\left| m_k \right\rangle$ are unique. In practice, this is typically a valid assumption.

Just before the aperture, the density matrix becomes

$$\rho_A = \sum_k s_k \Im \left| m_k \right\rangle \left\langle m_k \right| \Im^{-1}. \qquad (21)$$

The matrix element $\rho_A(x, x^{'})$ can be written as

$$\rho(x, x^{'}) = \sum_k s_k \left\langle x | \Im | m_k \right\rangle \left\langle m_k | \Im^{-1} | x^{'} \right\rangle. \qquad (22)$$

If we define $\Im \left| m_k \right\rangle = \left| M_k \right\rangle$ and $\left\langle x | M_k \right\rangle = M_k(x)$, then $M_k$ is the Fourier transform of $m_k$, and the matrix element takes on the simple form

$$\rho_A(x, x^{'}) = \sum_k s_k M_k(x) M_k^*(x^{'}). \qquad (23)$$

This expression can now be substituted into equation (18). Solving for $M_k(x)$ will allow us to retrieve the mode functions $m_k(y)$ via a simple inverse Fourier transform, where $y$ are the transverse coordinates in the source plane.

As an aside, we can define the purity of the source as $\wp(\rho_s) = Tr\left[\rho_S^2\right]$, which is easily calculated as

$$\wp(\rho_s) = Tr\left[\rho_S^2\right] = \sum_k s_k^2 \leq 1. \tag{24}$$

This is a useful figure of merit for single-mode behaviour of the source. The smaller $\wp$, the more mixed the source. This is a standard quantity in quantum information theory [23].

In what follows, we will calculate the $M_k(x)$ from the experimental intensities $I_j(u)$ that we measure. Ptychography is an exceptionally powerful solution of the Fourier domain phase problem [24], which allows us to solve for both the amplitudes of the modes and their phases. The iterative reconstruction algorithm we employ is the widely-used ePIE [19]. At a given point during the iterative procedure, we have a running estimate of the object function and the illumination function (which in this case will be composed of multiple modes within the aperture). By computationally propagating the corresponding exit waves (one for each mode) to the detector plane, we obtain a set of estimated complex wavefunctions incident upon the detector, $\Phi_{j,k}(u)$, for a particular $k^{th}$ mode and $j^{th}$ specimen position. In general, the sum of the intensity of these estimated modes will not correspond to the measured intensity of our experimental data (equation 18), i.e. $I_j(u) \neq \sum_k \left|\Phi_{j,k}(u)\right|^2$. We therefore calculate new estimates of the complex modes incident upon the detector

$$\Phi'_{j,k}(u) = \frac{\Phi_{j,k}(u)\sqrt{I_j(u)}}{\sqrt{\sum_{k=1}^{k}\left|\Phi_{j,k}(u)\right|^2}}, \tag{25}$$

where $I_j(u)$ is the intensity component arising on the detector at the $j^{th}$ position [5]. This constraint has the effect of scaling the intensity of each mode by a scalling factor so that their sum matches the measured intensity, while preserving their estimated (and immeasurable) phases. Next, each mode is back-propagated to object/aperture plane, where the object and each mode estimate are altered to be more closely commensurate with the Fourier constraint according to an update procedure, as shown in the flow diagram in Figure 2. The update function in real space is detailed for a single mode in [25]: the same update is in this case applied to all modes separately, but the object function is kept constant for all the updates. The process is repeated over all specimen positions many (typically a few hundred) times until an error metric, characterising the difference between the estimated detected intensity and the measured intensity reaches a certain threshold (see section 3 below); convergence properties are discussed in [25].

Diversity in the data, and the fact that the modes are added in intensity, ensures that each mode is independent (in the sense of being incoherent) with respect to every other mode. The modified Fourier constraint gives us $K$ modes, where $K$ is the number of illumination modes we have chosen to include in the reconstruction. There is an infinite combination of modes that can satisfy the overall intensity constraint, because they can be made up of any linear combination of each other [26].

However, the most compact description of the partially coherent state of the illumination is the set of orthogonal modes corresponding to the diagonalised density matrix (equation 20) and their corresponding illumination modes $M_k(x)$, of which there may be only $\kappa$ pure states with significant probability, with $\kappa < K$. Indeed, if the experiment is perfectly coherent we have by definition $\kappa = 1$. If the reconstruction uses $K < \kappa$, then both the object and the mode reconstructions are imperfect, leading to a higher error metric: the partial coherence has not been fully accounted for. However, using a large $K$ incurs a high computational overhead. The $M_k(x)$ give us the most compact representation of the partial coherence in the illumination, which is what we measure here.

It is important to note that the sum of the intensities of the source modes, $m_k(y)$, is not the same as the physical intensity distribution of the source, despite them being the back Fourier transform of the $M_k(x)$. The modes derive from the entire experiment, i.e. the information that arrives at the detector. A full description of the source requires us to capture the entire wavefunction, including the part that falls outside the aperture. Or, equivalently, the source we reconstruct is the real source as seen backwards from detector, which is an image that is diffraction limited by the aperture.

## 3. EXPERIMENT

Provided position-dependent aberrations in the image forming optics (coma, etc.) can be ignored and the specimen is suitably thin, the idealised experiment of Figure 1.a can be realised in practice using the TEM geometry shown in Figure 1.b; the projection of the selected area aperture back through the objective lens gives a virtual aperture in the specimen plane equivalent to the pre-specimen aperture in Figure 1.a.

We implement the setup of Figure 1.b using 300keV electrons in a JEOL R005 high-resolution transmission electron microscope (TEM) equipped with a cold field emission gun and dual aberration correctors. A condenser aperture with a diameter of 100μm was inserted to ensure a reasonable coherence of the electron wavefield illuminating the specimen, which was a thin layer of gold evaporated on amorphous carbon. The energy spread of the electron wave is small ($\Delta E/E < 10^5$), meaning that the path length differences (maximum $10^{-7}$ m given the geometry of the experiment) between any of the interference phenomena we measure are much smaller than longitudinal coherence length ($10^{-6}$ m).

Diffraction patterns were collected in the selected area diffraction mode at a calibrated camera length of 2.44m, using a 10μm selected area aperture. Usually in this mode a beam stop is needed to prevent the intense central spot of the diffraction pattern from saturating or damaging the detector, resulting in loss of the low spatial frequency information that this bright peak contains. To combat this here, instead of imaging the diffraction pattern in the far-field Fourier domain, as would be conventional, we adjusted the microscope intermediate lens to image a plane at a slight defocus from the microscope's primary image plane, resulting in near-field diffraction patterns whose much lower dynamic range removes the need for a beamstop. Equations (9) and (15) therefore do not formally apply, but this is not fundamental: any unitary operator (whether Fresnel or Fourier) can be used to propagate the $M_k(x)$ to the detector plane. In the reconstruction process we use the angular spectrum method [10] to calculate the near-field diffraction pattern before applying the intensity constraint in equation (25). For an example diffraction pattern, see Figure 3.

Diffraction patterns were recorded using a Gatan Orius SC2003 CCD camera with 2048×2048 pixels on a 7.4µm pitch. The detector was binned by 2 during the acquisition and the central 512 by 512 region of each pattern extracted. The specimen was placed on a holder driven by a piezo motor with a minimum incremental step of 0.3 nm and was translated through a 15×15 grid of positions in a raster fashion, with a diffraction pattern captured by the CCD at each of the 225 specimen positions. When projected back to the specimen plane the selected area aperture covered a disc of 113nm on the specimen surface, so the position grid was programmed with a nominal step size of 30 nm to ensure significant overlap of the selected specimen areas between positions. The stage repeatability was determined to be much better than its absolute accuracy, so we obtained a reasonable initial estimate of the true position grid by removing the selected area aperture and recording a set of defocussed brightfield images with the same moving instructions as for the subsequent ptychography experiments. These images were then cross-correlated to give a set of relative offsets for use as initial grid positions for the ptychographic algorithms, which further improved these initial position estimates using a simulated annealing method [25].

Four sets of diffraction patterns were used for the results presented here: dataset 1 and dataset 2 were collected at spot size 2 and spot size 4 respectively; dataset 3 was collected at a stronger strength of the second condenser lens at spot size 3, while dataset 4 was collected at a weaker strength of the second condenser lens at spot size 3. The spot size, controlled by the condenser lenses, determines the angular size of the source as seen from the specimen and aperture, and hence the spatial coherence of the illumination. The exposure time was varied between each set of data to make sure the highest pixel value in any given diffraction pattern was within the linearity range of the detector. The longest exposure was 3 seconds (dataset 2) and the shortest was 1 second (dataset 1). Due to slowly varying diffraction lens current instability and a relatively long data collection time, the centre of the diffraction patterns drifted during our experiments by around 50 pixels – this drift was corrected algorithmically during the reconstruction process, as detailed in [27].

The 'ePIE' implementation of the multi-mode ptychography algorithm [26] was used to process our data sets. However, for the first 500 iterations we used an approximate method for accounting for the partial coherence using a Gaussian convolution of the diffraction pattern, as described elsewhere [27,28]. With reference to Figure 4, which shows the error metric as a function of iteration number, we see that when the multi-mode method is switched on, the error drops significantly after an initial instability, indicating that the modal decomposition is much more effective at handling the coherence properties of the beam. We use $K=16$ illumination modes and a single object mode. It is computationally impractical to diagonalise the density matrix itself in the aperture plane (composed of $512 \times 512$ pixels), so after the reconstruction process is complete, we use an equivalent orthogonalisation method derived from principle component analysis. The modes are put into an $X$ by $K$ matrix, where $X$ is the number of pixels in the aperture plane. We form the $K \times K$ covariance matrix of the modes, and then find its eigenvalues and eigenvectors. The $K$ orthogonalised modes are then obtained by projecting the original modes onto the eigenvectors of the covariance matrix [26].

Figure 5 shows the first six components of the four partially coherent electron wavefields. The pixel sizes in the reconstructions are calculated according to

$$\delta u = \frac{\delta x \times D}{M_O \times d} \ , \tag{26}$$

where $\delta x$ is the pixel size of the detector after binning; $M_o$ is the magnification of the objective lens; $D$ is the physical size of the aperture and $d$ is the physical size of the central diffraction patterns disk. This gives pixel dimensions of 0.54nm, 0.48nm, 0.37nm and 0.38nm for datasets 1-4 respectively, and explains the discrepancy in the apparent size of the apertures shown in Figure 5. Figure 6 plots the eigenvalues of the reconstructed modes representing the partially coherent electron wavefield, which illustrates that as we change the strength of the condenser lenses, the degree of partial coherence changes: as the first condenser lens is strengthened or the second condenser lens is weakened , the effective size of the source is smaller, which results in better transverse coherence. Figure 7 compares ptychographic reconstructions from the spot 2 dataset using the multi-mode method (with Figure 7.a, showing the phase and 7.b showing the amplitude of the reconstruction) versus a conventional single mode reconstruction (with Figure 7.c showing the phase and 7.d showing the amplitude). The phase contrast and resolution of the multi-mode specimen reconstruction are clearly improved over the single mode version. The effective source functions finally resulted from these reconstruction (i.e. the intensity summation of the Fourier transform of the 16 modes) are clearly not symmetric – an assumption usually made in the electron microscopy literature. The variation in their size corresponds to the change of magnification of the source as the condenser lenses are varied, just as we would expect.

## 4. DISCUSSION

The degree of partial coherence in propagating electron waves has previously been inferred by measuring the transverse mutual intensity function via holography or other interferometric methods, usually only in one dimension. In electron coherent diffractive imaging, another approach to handle the spatial coherence envelope is to model it as a Gaussian function of the illumination convergence angle that dampens the transfer of high spatial frequencies through the TEM [7]. However, this Gaussian model assumes the profile of the source is symmetric; the results shown here clearly show that in general it is not.

Here we use a comprehensive method to model partial coherence, which combines ptychography with modal decomposition to measure the components of a partially coherent electron wavefield explicitly, without any assumptions about the source shape or its spatial coherence properties. Uniquely amongst the various forms of coherent diffractive imaging, the redundancy and diversity of the ptychographic dataset is capable of supplying the greater amount of information required to retrieve multiple modes from diffraction data. The particular representation of the modes we have calculated, corresponding to the pure states associated with the eigenvalues of density matrix, is the most comprehensive and compact description of the partial coherence in a propagating matter wave. We emphasise that the wavefield reconstructed from our data is not a representation of the actual electron source, but rather a diffraction limited image of source as it is seen through the aperture from the detector; the effective sources shown in Figure 5 are the convolution of the source function with the Fourier transform of the aperture function.

The formal theoretical analysis we derive explicitly includes the spatial coherence of the source, but can be expanded in a straightforward manner to include other effects. For example, our source has chromatic spread, the lenses employed may have high frequency instabilities and the specimen will have generated inelastic scattering events that may partly mask the coherent (elastic) scattering signal. We have also not accounted for the pixel size and fill factor of the detector. These effects will have to be the subject of further work. However，in our experience to date with visible light and X-ray ptychography, other sources of incoherence usually manifest themselves as modes that bear no relation to the size or shape of the illumination.

The partial coherence of the source is a key obstacle to accurate phase retrieval imaging. Ptychography combined with multi-mode reconstruction techniques offers a highly flexible way forward, since it considers the multiple sources of partial coherence within a single framework that does not require *a priori* coherence measurements or a parameterisation of the coherence envelope based on simplifying assumptions. The goal of our future research is therefore to optimise our experimental process, maximising information content in the diffraction data to realise high resolution electron phase imaging whilst retaining the accuracy and sensitivity already demonstrated at lower resolutions.

## ACKNOWLEDGEMENTS

This work was funded by Phase Focus Ltd, EPSRC Grant No. EP/1022562/1 and the Department of Electronic and Electrical Engineering, University of Sheffield. JMR declares a shareholding in Phase Focus Ltd, which owns patents relating to some of the methods described in this paper.

**Figure Captions**

Figure 1.a the idealised configuration. The upper part includes the source and the aperture which lies in the far field corresponding to the source; the lower part is the detector lies in the Fourier domain of the aperture.

Figure 1.b the setup of the experiment. The specimen was shifted by the piezo stage in x and y directions. The selected area aperture was on the first image plane with a slight defocus. Data 1 and data 2 were collected at spot size 2 and spot size 4, which can be taken as if the source were at different heights; the higher one is the spot 4. Data 3 and data 4 were collected at a different convergence angle by changing the strength of the condenser lens.

Figure 2. The flow chart when multiple probe modes engaged in ePIE. N modes were reconstructed after applying the calculation in the frame at every position for several hundreds of iterations. Then the N reconstructed probes (mixed states) will be orthogonalised to give the eigen modes.

Figure 3. One of the raw experimental diffraction patterns, which were collected in the near field (see text for more information.).

Figure 4. The error metric of data 4 with 500 iterations Gaussian convolution partial coherence correction and 800 iterations multiple modes partial coherence correction. The error at $500^{th}$ iteration $e_G = 0.003833$; the final error $e_m = 0.002802$. After 500 iterations Gaussian corrector then from the $501^{st}$ iteration modes corrector began without updating the object until the 901st iteration.

Figure 5. The modes and the effective sources of the 4 sets of data. In any set of modes, each of them is orthogonal to each other. The numbers above each mode show the contribution percentage of each mode. The effective sources are the convolution of the source with the aperture.

Figure 6. The plots of the 16 eigenvalues of the 4 data sets. The more coherent the source is, the more contribution comes from the first mode, and less come from the other modes.

Figure 7. The specimen reconstructions. (a) and (b) show the phase and modulus of the specimen when 16 modes were engaged. (c) and (d) show the phase and modulus reconstructed with single mode. The main images are the magnification of the squared area by the red dashed line in the whole field of view inserted in the top corner.

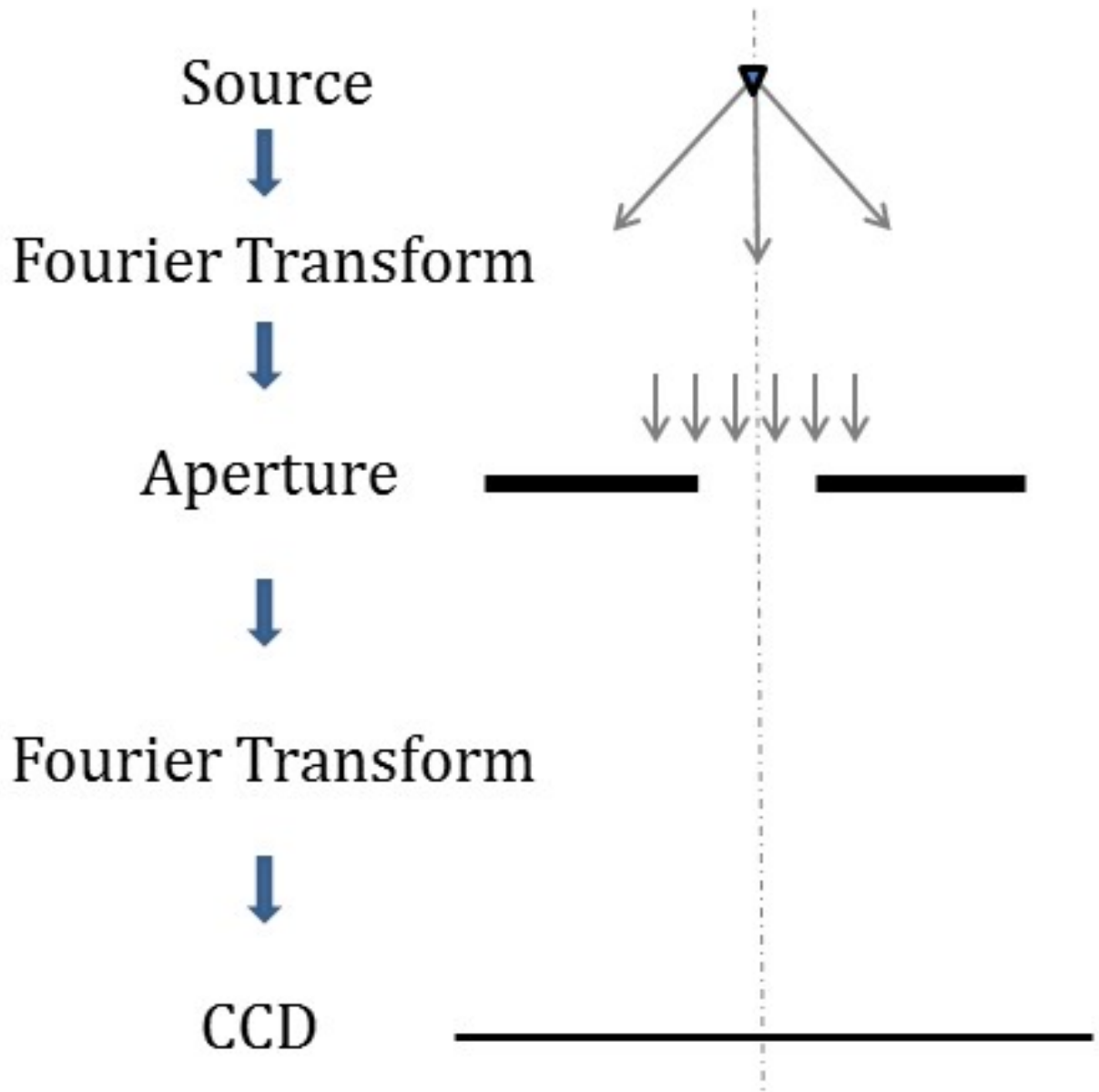

Figure 1.a

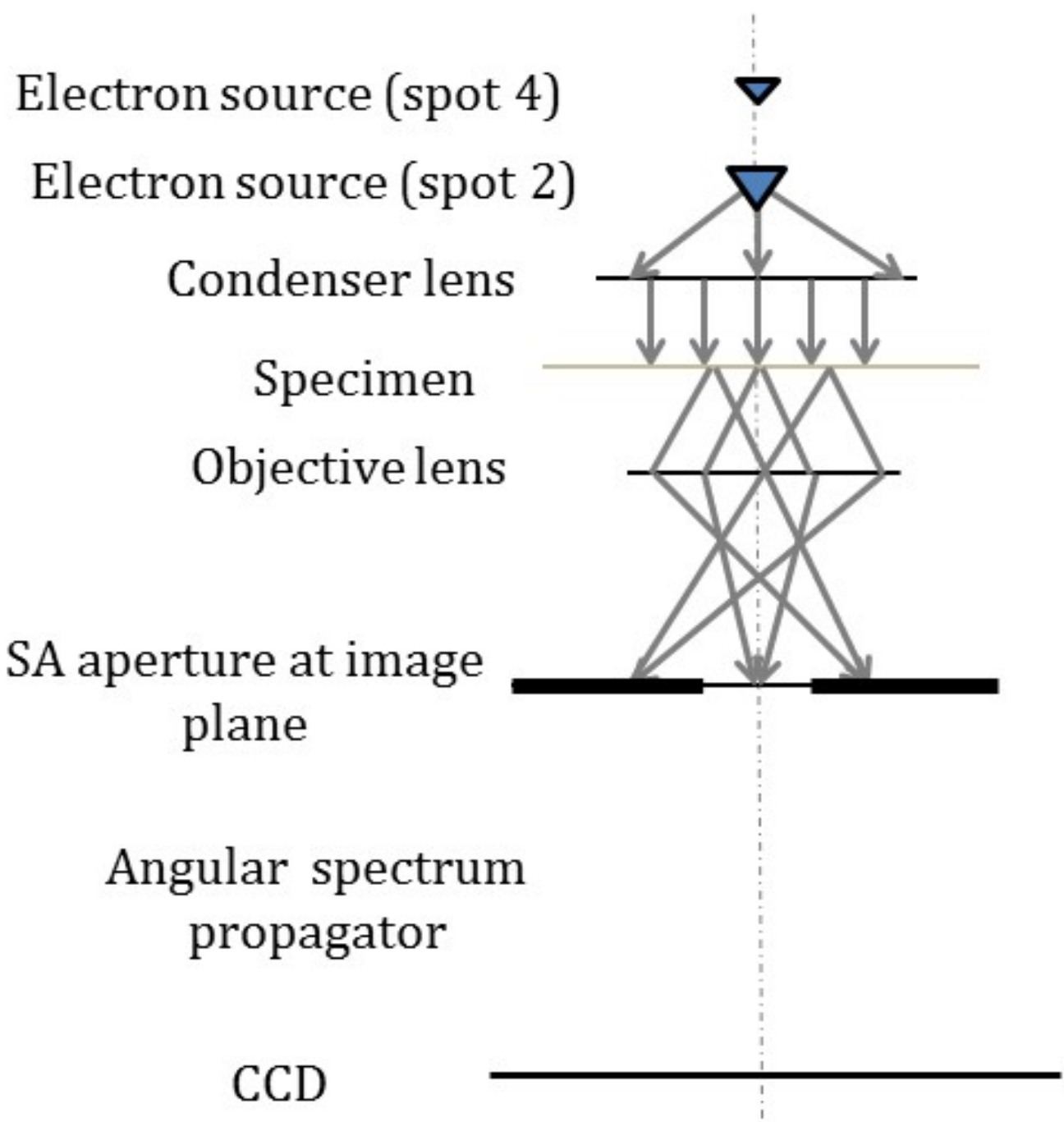

Figure 1.b

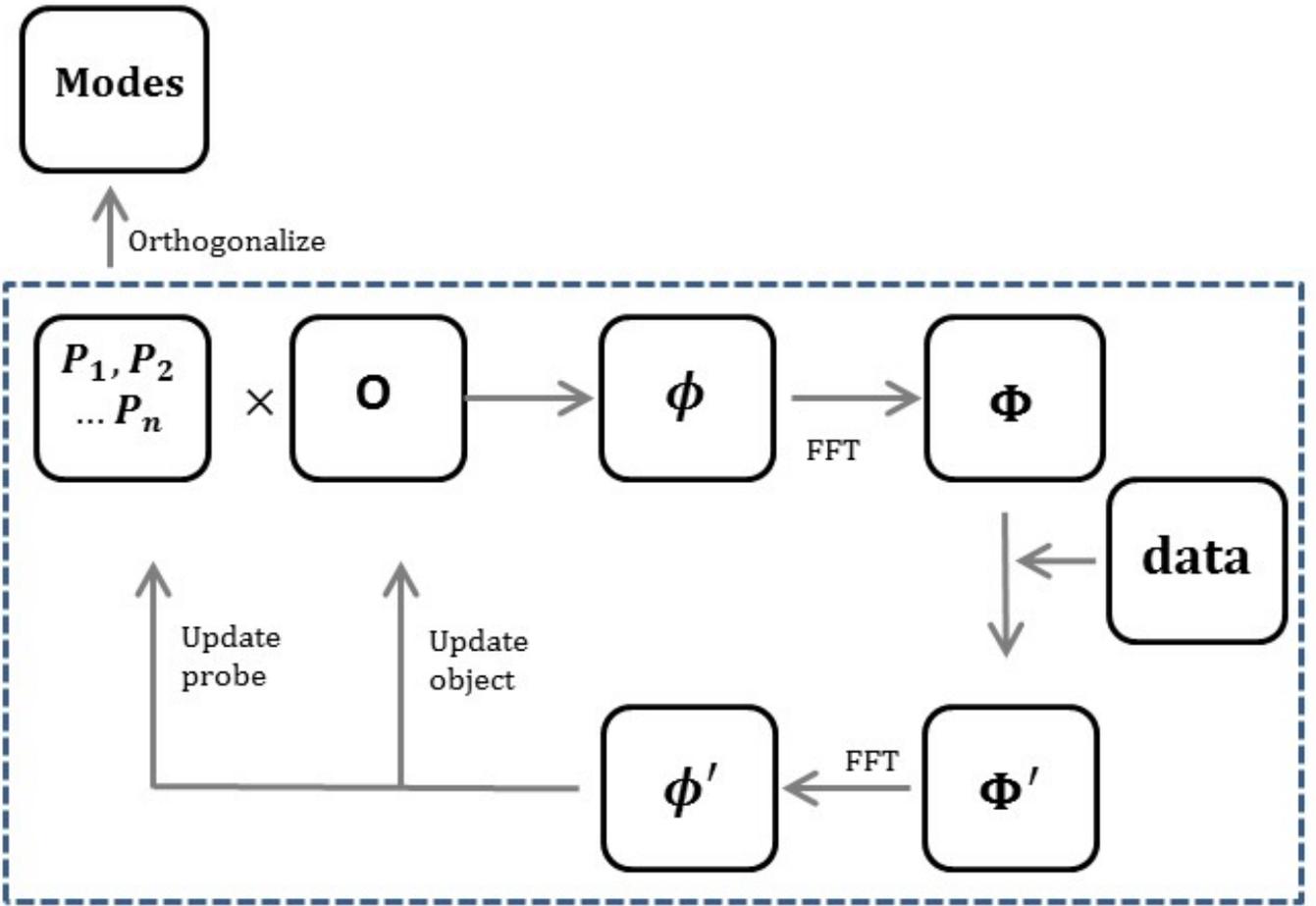

Figure 2

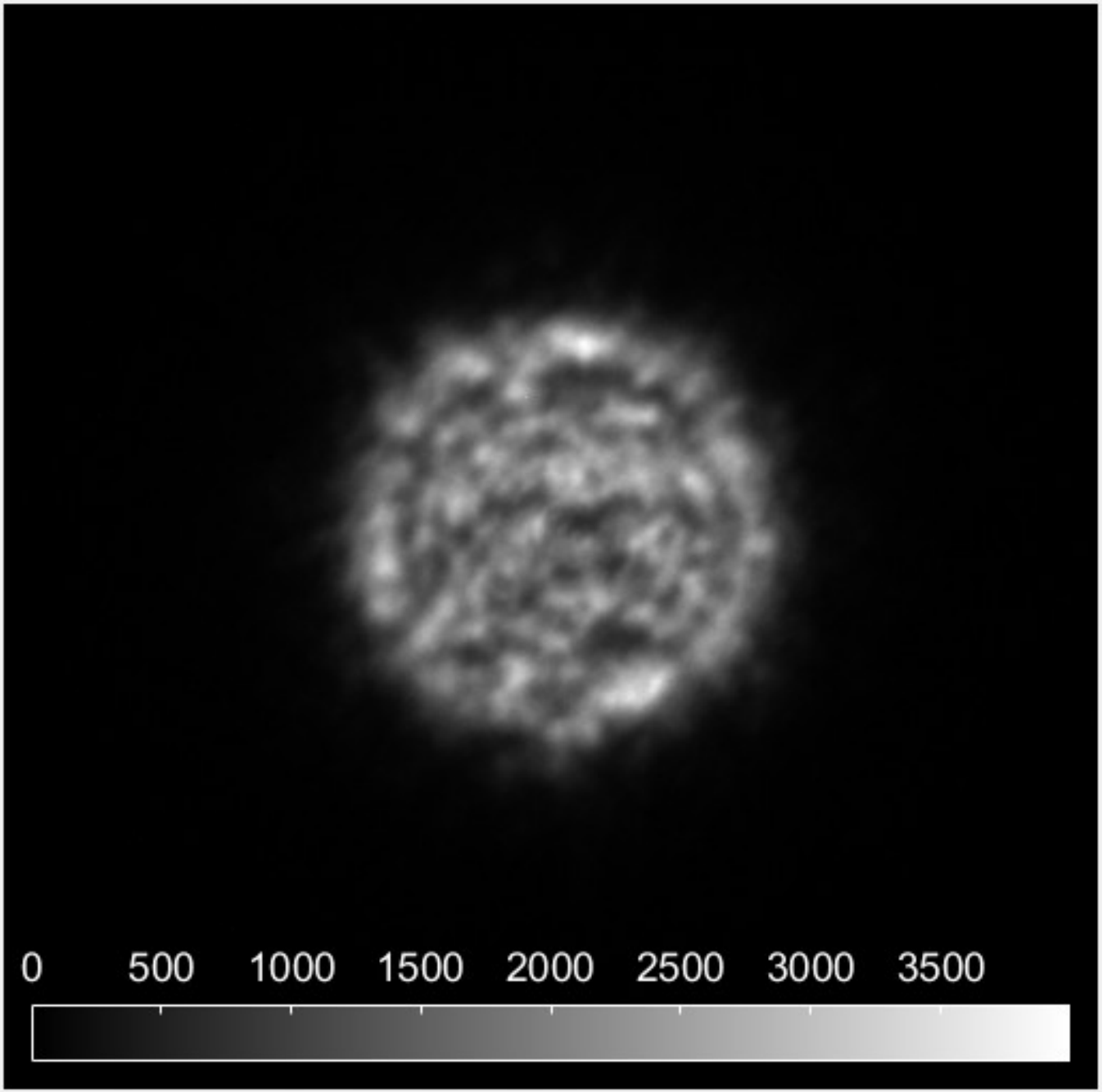

Figure 3

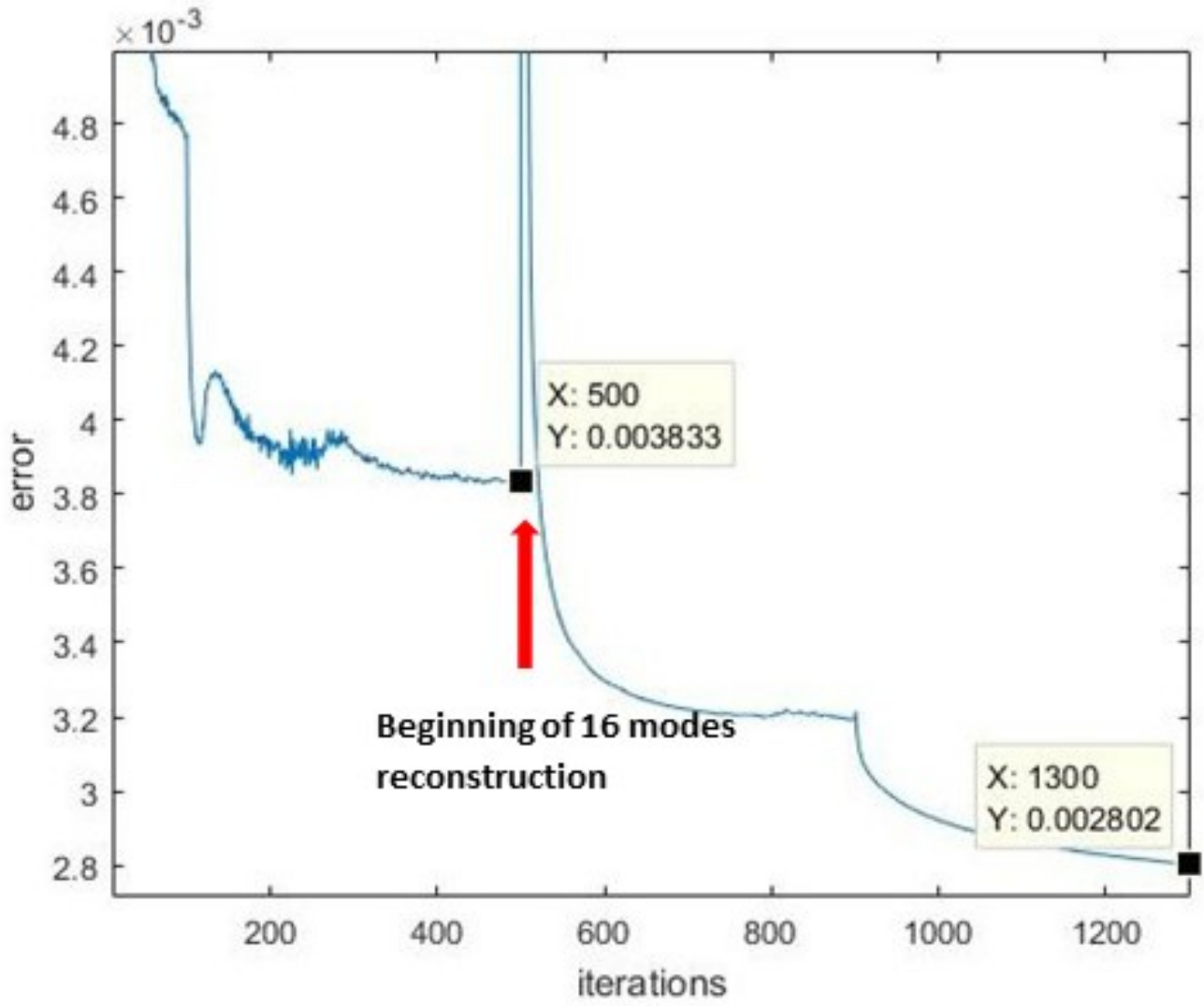

Figure 4

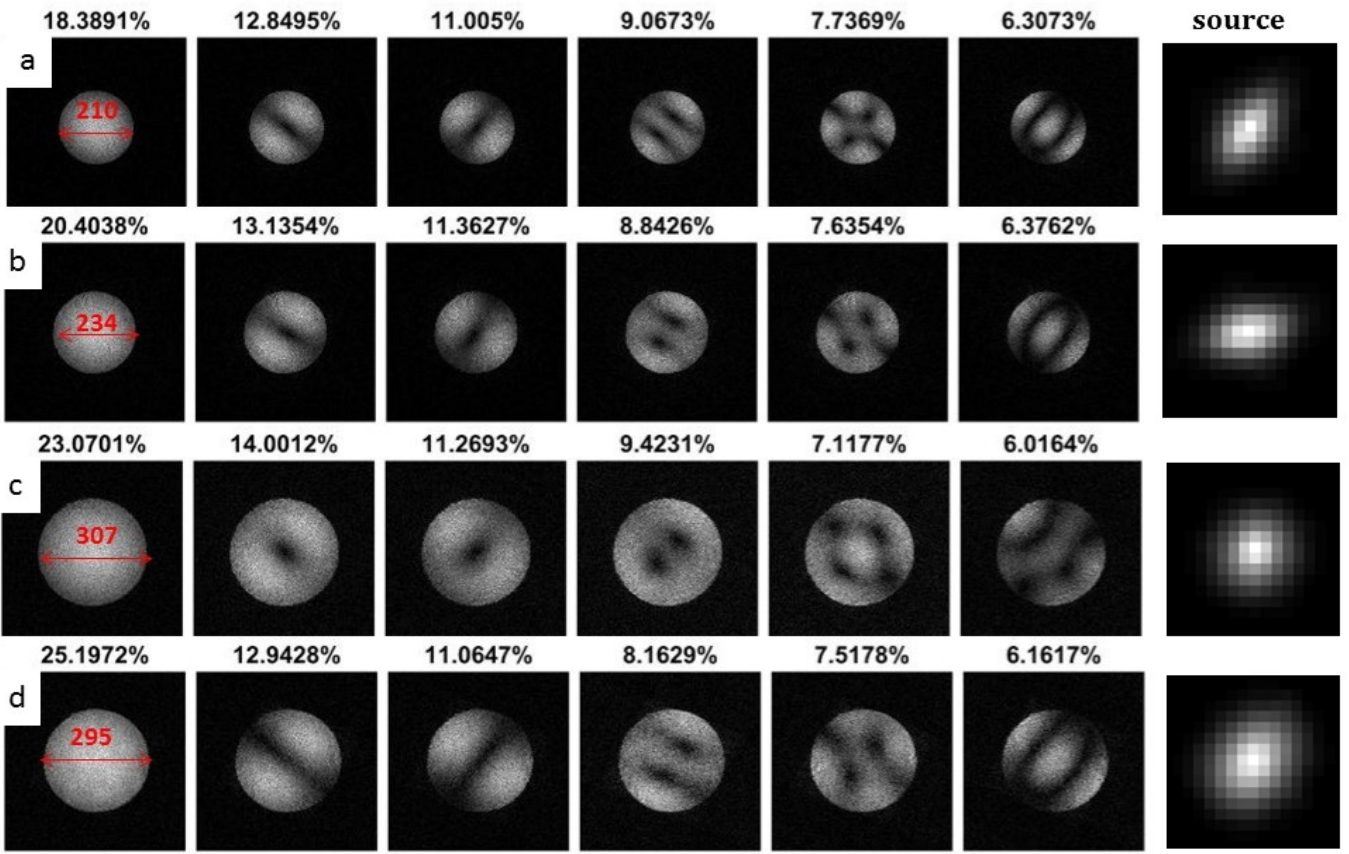

Figure 5

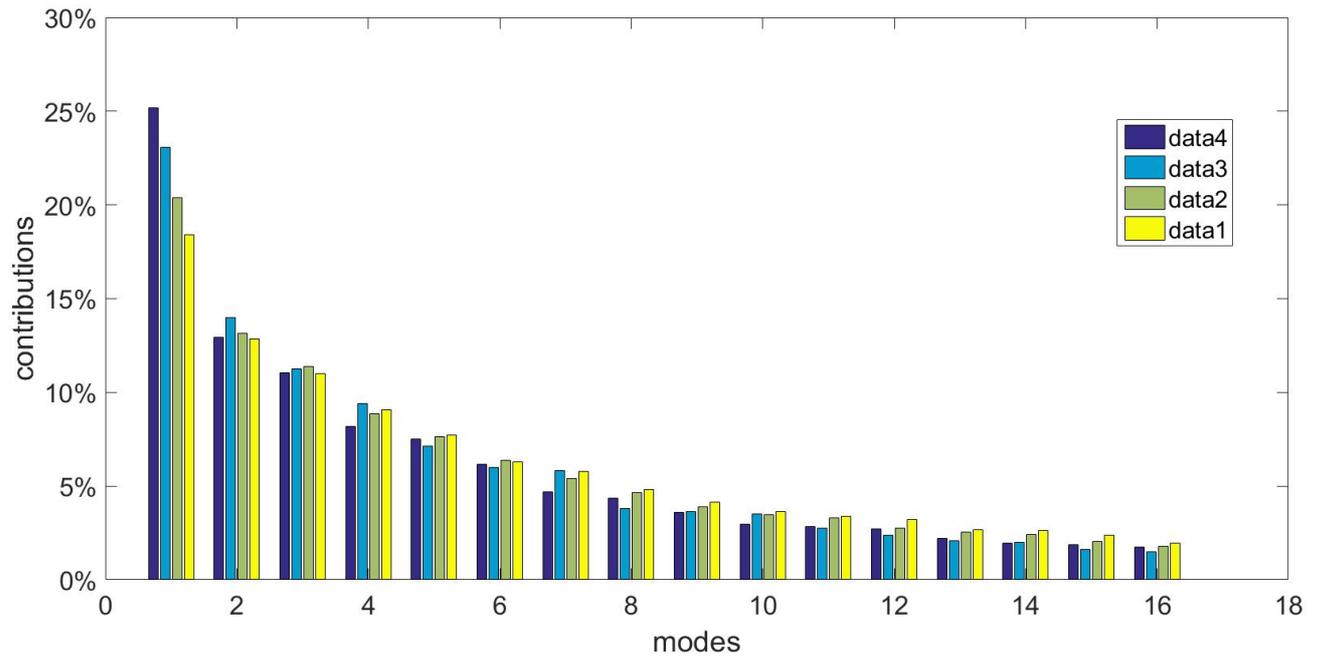

Figure 6

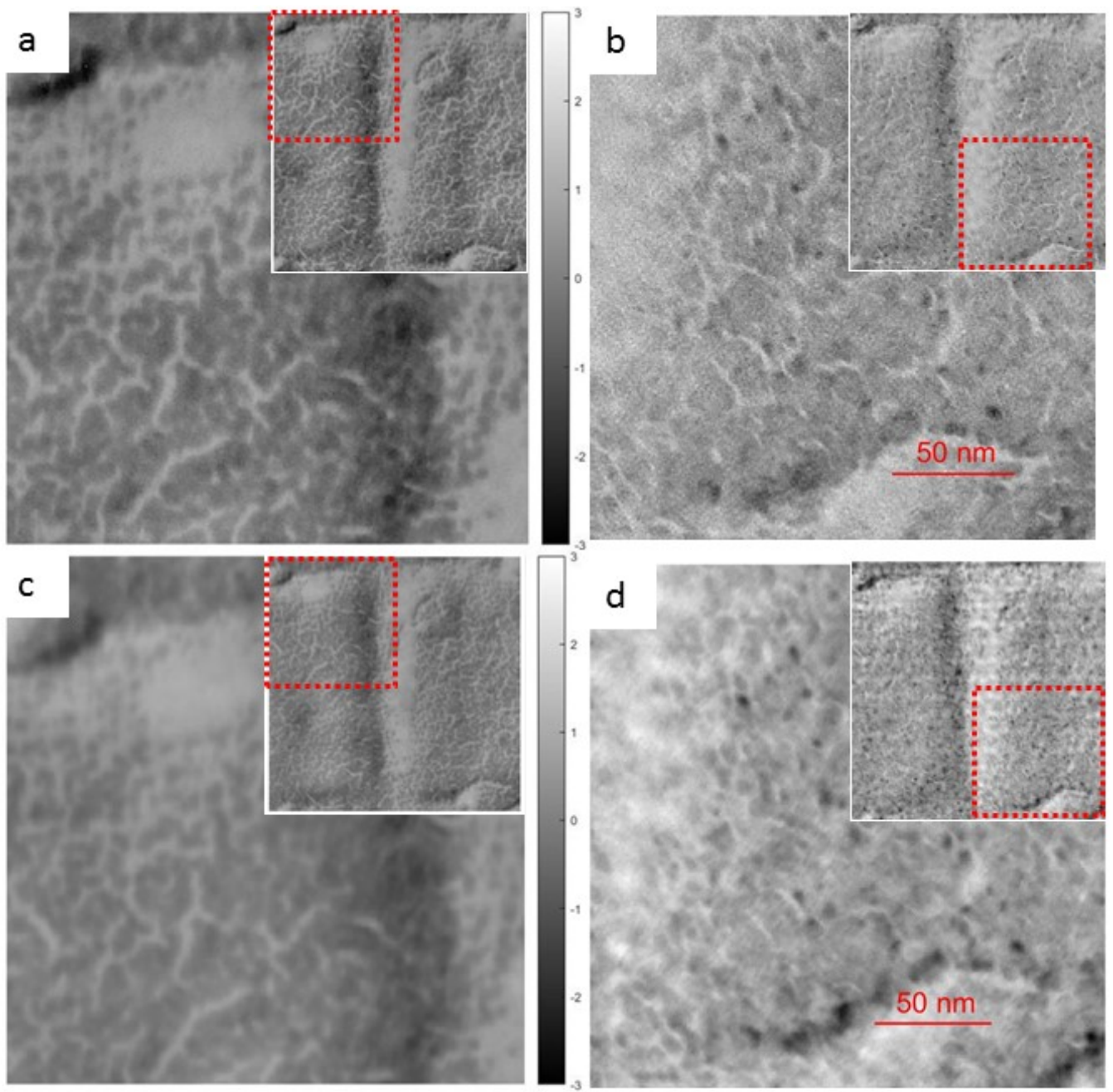

Figure 7